\documentclass[aps,letterpaper,twocolumn,preprintnumbers,floatfix,superscriptaddress]{revtex4}
\usepackage[colorlinks,linkcolor=black,anchorcolor=black,citecolor=black]{hyperref}
\usepackage{dsfont}
\usepackage{graphicx}
\usepackage{dcolumn}
\usepackage{bm}
\usepackage{epstopdf}
\usepackage{epsfig}
\usepackage{gensymb}
\usepackage{mathrsfs}
\usepackage{amsmath}
\usepackage{amssymb}
\usepackage{graphicx,bm}
\usepackage{slashed}
\usepackage[makeroom]{cancel}
\usepackage{color}
 \usepackage[titletoc]{appendix}

\def\fun#1#2{\lower3.6pt\vbox{\baselineskip0pt\lineskip.9pt
  \ialign{$\mathsurround=0pt#1\hfil##\hfil$\crcr#2\crcr\sim\crcr}}}

\def\lsim{\mathrel{\rlap{\raise 2.5pt \hbox{$<$}}\lower 2.5pt\hbox{$\sim$}}}
\def\gsim{\mathrel{\rlap{\raise 2.5pt \hbox{$>$}}\lower 2.5pt\hbox{$\sim$}}}

\input epsf

\newcommand{\be}{\begin{align}}
\newcommand{\ee}{\end{align}}
\newcommand{\bea}{\begin{eqnarray}}
\newcommand{\eea}{\end{eqnarray}}

\begin{document}

\title{Generating a Higgs Quartic}

\author{Csaba Cs\'aki}
\affiliation{ Department of Physics, LEPP, Cornell University, Ithaca, NY 14853, USA}

\author{Cong-Sen Guan}
\affiliation{ Department of Physics, LEPP, Cornell University, Ithaca, NY 14853, USA}
\affiliation{
 CAS Key Laboratory of Theoretical Physics, Institute of Theoretical Physics,
Chinese Academy of Sciences, Beijing 100190, China.}
\affiliation{School of Physical Sciences, University of Chinese Academy of Sciences, Beijing 100049, P. R. China.}

\author{Teng Ma}
\affiliation{
 CAS Key Laboratory of Theoretical Physics, Institute of Theoretical Physics,
Chinese Academy of Sciences, Beijing 100190, China.}
\author{Jing Shu}
\affiliation{
CAS Key Laboratory of Theoretical Physics, Institute of Theoretical Physics,
Chinese Academy of Sciences, Beijing 100190, China.}
\affiliation{School of Physical Sciences, University of Chinese Academy of Sciences, Beijing 100049, P. R. China.}
\affiliation{CAS Center for Excellence in Particle Physics, Beijing 100049, China}
\affiliation{Center for High Energy Physics, Peking University, Beijing 100871, China}

\begin{abstract}
We present a simple mechanism for generating a Higgs quartic in composite Higgs models without a corresponding quadratic term. The extra quartic will originate from a Higgs dependent kinetic mixing between additional fermionic states. The mechanism can be naturally embedded to models with maximal symmetry as well as Twin Higgs models.  The resulting Twin Higgs models will have a fully natural realistic Higgs potential, where the quartic mechanism will serve as the only source for the $Z_2$ breaking, while the top and gauge sectors can remain exactly $Z_2$ invariant.

\end{abstract}


\maketitle

\section{Introduction}
\label{Sec:intro}
The sensitivity of the Higgs potential to very large energy scales makes it unstable to quantum corrections. Stabilization of the scale of electroweak symmetry breaking (EWSB) is
one of the most profound problems in particle physics. One can eliminate this sensitivity to UV scales by introducing new physics (such as supersymmetry) not too far above the EWSB scale. Of such new physics models, the pseudo-Nambu-Goldstone (pNGB) Higgs~\cite{Kaplan:1984plb,Georgi:1984plb,Dugan:1985npb,ArkaniHamed:2002qy,Agashe:2004rs} (for reviews see~\cite{Contino:2010rs,Bellazzini:2014yua,Panico:2015jxa,Schmaltz:2005ky}) is one of the simplest and most widely studied. In such models the Higgs potential originates from the interactions that explicitly break a shift symmetry that would otherwise forbid the generation of a Higgs potential. The mechanisms that ensure that the resulting Higgs potential remains free of UV divergences
include collective symmetry breaking~\cite{ArkaniHamed:2002qy}, discrete symmetry~\cite{Chacko:2005pe,Csaki:2017jby} or maximal symmetry~\cite{Csaki:2017cep,Csaki:2018zzf}.

 The main difficulty of such pNGB Higgs models is that   the Higgs quadratic and quartic terms are
generically strongly correlated  making it very difficult to separate the scale of new physics from the scale of EWSB, which conflicts with precision measurement and direct detection. Since the quadratic and quartic terms are generated by the same dynamics, it is usually hard to enhance the quartic without also increasing the quadratic term. Hence usually a tuning of order a few percent level is needed to produce the little hierarchy of the EWSB and the new physics scales.
Models that produce an adjustable Higgs quartic term without introducing a Higgs quadratic provide an elegant solution to the little hierarchy problem. An example of this type are 6D models~\cite{Csaki:2017eio}
where a tree-level quartic can originate from the gauge boson components along the extra dimension, or the little Higgs models~\cite{ArkaniHamed:2001nc, ArkaniHamed:2002qy} based on dimensional deconstruction of the 6D theory.
However these models are usually quite complicated and also require additional pNGBs, such as the second Higgs doublet (ie. the generated quartic is that of a two-Higgs doublet model and not a true SM-like Higgs quartic).

In this work we propose a novel mechanism to produce an adjustable Higgs quartic self-coupling from loop corrections without a corresponding Higgs quadratic term. We introduce an electroweak (EW) triplet and singlet fermion and observe that if their kinetic terms are independent of the Higgs field, we will only produce a Higgs quartic term in the 1-loop effective potential but no Higgs quadratic term. The simple underlying reason is that a triplet-singlet mixing neccessarily involves at least two Higgs insertions. Moreover, the sign of the generated quartic will be positive if the Yukawa term mixing the triplet and the singlet is momentum dependent (while momentum independent mixing term always give a negative contribution). The recently proposed maximal symmetry \cite{Csaki:2017cep,Csaki:2018zzf} has exactly the right properties for this mechanism: it's main effect is exactly to protect the effective kinetic terms from Higgs dependent corrections. Thus models with maximal symmetry can naturally produce a positive and adjustable Higgs quartic term. Our mechanism can be simply implemented in any pNGB Higgs model based on deconstruction or warped extra dimensions scenario without having to introduce any  additional structures. We show that the tuning in these models is greatly reduced, for example the minimal implementation of maximal symmetry will have about 5\% tuning. In twin Higgs models the additional quartic will allow the top and gauge sectors to remain exactly $Z_2$ invariant, leading to models with no tuning whatsoever.

The structure of this paper is organized as follows. In Sec.~\ref{Sec:brief_ideas} we explain our mechanism of generating a Higgs quartic term. In Sec.~\ref{sec:mixing} we show how the requisite Higgs-dependent kinetic mixing can be obtained from integrating out heavy fields.
In Sec.~\ref{Sec:details_mechanism} we present a concrete realization of this mechanism in the $SO(5)/SO(4)$ pNGB Higgs model based on the two site moose with minimal maximal symmetry\cite{Csaki:2018zzf}.
In Sec.~\ref{Sec:twin_Higgs} we apply our mechanism to the $SO(8)/SO(7)$ Twin Higgs model  which will allow the top and gauge sectors to remain exactly $Z_2$ symmetric. In Sec.~\ref{Sec:Fine_tune} we discuss EWSB and the fine tuning needed in the two example models and show some numerical results. We find that the colored top partners can be heavy enough to evade LHC direct detection with modest tuning in the first model, while fully natural EWSB without any tuning can be achieved in the Twin Higgs model. We conclude in~Sec. \ref{Sec:conclusion}. The appendices contain the detailed expressions of the form factors in the effective Lagrangian, the descriptions of the top and gauge sectors of model with maximal symmetry, as well as the details of the Twin Higgs model.

\section{Generation of the Higgs Quartic}\label{Sec:brief_ideas}

In this section we illuminate the essence of our simple mechanism for producing an adjustable Higgs quartic coupling based on pNGB Higgs model. We  introduce the electroweak (EW) triplet $\Delta$ and singlet $\eta$ Dirac fermions, both of which are assumed to be elementary. For simplicity first we assume that they are massless (but in the full model all allowed mass and Yukawa terms will be added). If the triplet and singlet mix through a Yukawa coupling involving the SM Higgs in the low energy effective theory, in momentum space the leading order Lagrangian will be given by
\bea
\mathcal{L} = \text{Tr}[ \bar{\Delta}  \slashed p \Delta] +\bar{\eta}  \slashed p \eta -\big(  \frac{\lambda}{f} H^\dagger \bar{\Delta} H \eta  + h.c.\big),
\label{eq:Yukawamixing}
\eea
where $f$ will be the pNGB Higgs decay constant and $H$ is the Higgs doublet. The quantum numbers of the two Dirac fermions under $SU(2)_L \times U(1)_Y$ are
\bea
\Delta \equiv
\frac{1}{\sqrt{2}} \left( \begin{array}{cc}
     \Delta^0 & \sqrt{2} \Delta^+ \\
     \sqrt{2} \Delta^- &  -\Delta^0\\
  \end{array} \right)  \in {\bf 3_0 }, \quad \eta \in {\bf 1_0}.
\eea
Note that the choice of triplet and singlet representations under $SU(2)_L$ is essential: since
$\Delta$ carries two $SU(2)$ indices and $\eta$ does not carry any, their Yukawa coupling has to contain at least two Higgs doublets.  Thus if this mixing terms is the only Higgs dependent term in the effective Lagrangian (\ref{eq:Yukawamixing}) while the kinetic terms are Higgs independent,
then one can only get a contribution to the Higgs quartic term without obtaining a shift in the Higgs mass term.

Treating the Higgs as a background field, we can obtain the leading one loop correction to the Higgs potential from the loop of the triplet and singlet fermions:
\bea \label{eq:Higgs_potential}
V(H) &\sim&  \frac{i}{2}  \int \frac{d^ 4 p }{(2\pi)^4} \frac{ \lambda^2 (H^\dagger H)^2}{f^2} \text{Tr}[\frac{i \slashed p }{p^2} \frac{i\slashed p }{p^2}]  \nonumber \\
&=& -  \frac{\lambda^2 (H^\dagger H)^2}{f^2}  \int \frac{d^ 4 p_E }{(2\pi)^4} \frac{2}{p_E^2},
\eea
where in the second line we performed a Wick rotation to Euclidean space  $p^2\to  -p_E^2$, where $p_E^2=p_0^2 +(\vec{p} )^2$ is positive definite. Thus we find that the Yukawa coupling of triplet and singlet fermions always produces a negative correction to the Higgs quartic coupling, unlike what one needs for successful EWSB.  From this detailed examination of the correction it is however clear how this problem can be solved: we expect that if one uses a Higgs dependent kinetic mixing rather than a Yukawa mixing the sign could be reversed, since we will pick up an extra $p^2$ term in the Feynman diagram, which after Wick rotation will provide an additional sign flip.

The Lagrangian describing the kinetic mixing of the triplet and singlet fermions  can be parametrized as
\bea
\mathcal{L} = \text{Tr}[ \bar{\Delta}  \slashed p \Delta] +\bar{\eta}  \slashed p \eta -\big(  \frac{\lambda}{f^2} H^\dagger \bar{\Delta} H \slashed p  \eta  + h.c.\big),
\eea
where we can see that the Yukawa coupling  now contains the extra momentum factor.
The leading one loop correction to the Higgs potential  can be explicitly expressed for this case as
 \bea
 V(H) &\sim&   \frac{i}{2} \int \frac{d^ 4 p }{(2\pi)^4}   \frac{ \lambda^2 (H^\dagger H)^2}{f^4} \text{Tr}[\frac{i(\slashed p )}{p^2} \slashed p    \frac{i(\slashed p )}{p^2} \slashed p]  \nonumber \\
 &=& \frac{2\lambda^2 (H^\dagger H)^2}{f^4}  \int \frac{d^ 4 p_E }{(2\pi)^4}.
 \eea
As expected an extra $p^2$ factor shows up with compared to the previous case, which will flip the sign of the contribution to the Higgs quartic after the Wick rotation to Euclidean space is performed. Thus we find that for the case of kinetic mixing the induced Higgs quartic is always positive.
We emphasize again that we have made the crucial assumption that the effective kinetic terms of the triplet and singlet fermions are Higgs in\-de\-pen\-dent. Otherwise corrections to the Higgs mass term will also be produced, and the quadratic and quartic terms would remain linked.
Note that for the case of scalar triplet and singlet both the Yukawa and kinetic mixings will always produce a negative shift in the Higgs quartic. For the case of mass mixing of scalars the propagators in the loop will contribute a factor of $i^2/p^4$ which will result in the same sign as the case of mass mixing with fermions after the Wick rotation is performed. For the case of scalar kinetic mixing we gain a factor of $p^4$ (rather than the $p^2$ for the case of kinetic fermionic mixing) thus the sign will not be flipped in this case. Hence we can see that only the case of kinetic fermionic mixing will produce the desired positive shift in the Higgs quartic self coupling.

To summarize we found two necessary conditions for producing an adjustable and positive Higgs quartic in the triplet-singlet model:
 \begin{itemize}
\item The effective kinetic terms must be Higgs independent;
\item The triplet-singlet mixing must be momentum dependent.
\end{itemize}
The first condition is exactly the main consequence of models with maximal symmetry, which we will take advantage of. In the next section we will briefly explain how the desired Higgs dependent kinetic mixing can be obtained.

\section{Generation of the Effective Kinetic Mixing\label{sec:mixing}}

So far we have established that a positive shift in the Higgs quartic can be obtained from a theory with a kinetic mixing between the triplet and singlet fermions. Before we present our full model we would like to first demonstrate how such a mixing can be easily generated in the effective theory. The key is to consider a chiral mixing between the fermions $\Delta , \eta$ and some heavy fermion $\psi$. Such linear couplings between elementary ($\Delta , \eta$) and composite ($\psi$) fields show up naturally in models of partial compositeness in composite Higgs models. For a simple illustration we introduce an $SU(2)_L$ doublet fermion  $\Psi_\mathbf{2}$ with the most general chiral mixing with the $\Delta ,\eta$:
\begin{equation}\label{eq:int_Lag}
  \mathcal{L}_{\text{int}}=\lambda_{1L}\bar{\Psi}_{2_R}\Delta_LH+\lambda_{2L}\bar{\Psi}_{2_R}H\eta_L+(L\leftrightarrow R)+h.c.
\end{equation}
After integrating out the heavy fermion we find the following effective mixing terms in the low-energy effective Lagrangian:
\begin{align}\label{eq:effct_mix}
  \mathcal{L}_\text{eff}^\text{mix}&=\frac{1}{M^2-p^2}\big(\lambda_{1L}\lambda_{2L}H^\dag\bar{\Delta}_LH\slashed{p}\eta_L\nonumber\\
  &+M\lambda_{1L}\lambda_{2R}H^\dag\bar{\Delta}_LH\eta_R\big)+(L\leftrightarrow R)+h.c.,
\end{align}
where $M$ is the mass of the heavy field. We can see that in the general case we get both the kinetic and Yukawa mixings in the effective Lagrangian (leading to both positive and negative contributions to the Higgs quartic). However one can easily turn off either the mass or the kinetic mixing by dialing the various $\lambda_{L,R}$ couplings.
For example, if we turn off the right handed or the left  handed couplings (e.g. $\lambda_{1,2R}=0$ or $\lambda_{1,2L}=0$), we will only get the kinetic mixing term, while if we turn off one left handed and one right handed coupling (e.g. $\lambda_{1R}=0$ and $\lambda_{2L}=0$), we will only get the Yukawa mixing term.

The simple lesson from this toy example is that a purely chiral mixing with the heavy fermion  will  produce the desired kinetic mixing in the effective theory. Below we will put all the various ingredients discussed above together to produce a realistic model realizing the mechanism  for an enhanced quartic.

\section{Implementation in the Simplest Two Site Model}\label{Sec:details_mechanism}

So far we have explained what the essential ingredients needed for generating the positive shift in the quartic coupling are. In this section we will show how to actually obtain a complete realistic model of this sort by embedding it into a simple 2-site composite Higgs model. For this we will use the simplest implementation of maximal symmetry recently proposed in~\cite{Csaki:2018zzf}. As explained above we need to use maximal symmetry in order to ensure that the kinetic terms of the fermions do not themselves depend on the Higgs field (preventing the generation of a shift in the mass term). First we review the construction of the minimal model with maximal symmetry and then add the additional fields needed to implement the mechanism involving the Higgs dependent triplet singlet kinetic mixing.

\begin{figure}
  \centering
  \includegraphics[width=7cm]{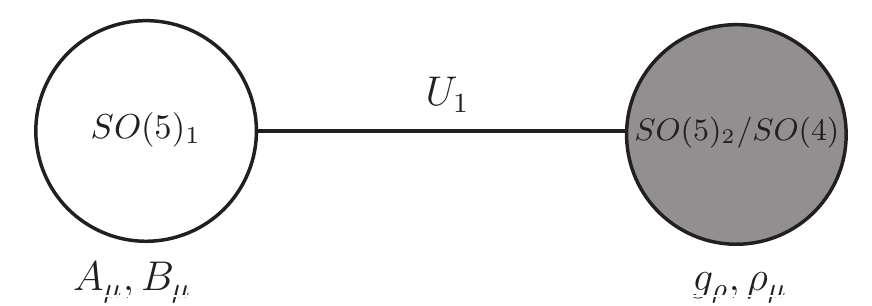}
  \caption{Gauge sector of the two-site model}\label{fig:2site_moose}
\end{figure}

The minimal model with maximal symmetry is using two sites which realized the  $SO(5)/SO(4)$ coset space. It can easily be generalized to N sites or a full continuous extra dimension but in this paper we will only focus on the simplest two site version.   Both sites will have  an $SO(5)$ global  symmetry thus the full global symmetry  is $SO(5)_1 \times SO(5)_2$.  A link field $U_1$  in the bi-fundamental representation of the global symmetry connects these two sites and breaks the global symmetry to the diagonal subgroup $SO(5)_V$.  The $SU(2)_L \times U(1)_Y$ subgroup of $SO(5)_1$ at the first site is gauged and identified with the usual  EW symmetry, while at the second site the entire $SO(5)_2$ is gauged, as shown in Fig.~\ref{fig:2site_moose}. This gauged $SO(5)_2$ is critical for the appearance of maximal symmetry. To realize the $SO(5)/SO(4)$ coset, the gauge symmetry at the last site should be broken to $SO(4)$ via a scalar in the 5-dimensional vector representation of $SO(5)_2$ with VEV $\mathcal{V} =(0,0,0,0,1)$. The linearly realized  pNGB field $\mathcal{H}'$ corresponding to this symmetry breaking can be parametrized as $\mathcal{H}' =U^{\prime} \mathcal{V}$, where $U^\prime$ is the non-linear sigma field of the coset space correponding to the breaking on the second site $SO(5)_2/SO(4)$. Some of the pNGB's will be eaten by the gauge bosons that become massive. In the end we are left with a single set of pNGBs corresponding to the $SO(5)/SO(4)$ coset. These uneaten pNGBs can be described by the linear sigma field $\mathcal{H} =U\mathcal{V}$ in the fundamental representation of $SO(5)_1$ with $U=U_1 U^\prime$.  Under unitary gauge, only the physical Higgs $h$ remains, and the field $\mathcal{H}$ can be parametrized as
\bea
\mathcal{H} =(0,0,0,s_h, c_h),
\eea
with $s_h \equiv \sin \big(h/f \big)$ and $c_h \equiv \cos \big(h/f \big)$.

Once we have the setup for the coset space we introduce the fermions in a way that corresponds to the minimal implementation of maximal symmetry. For this
the triplet $\Delta$ should live on the first site and the singlet $\eta$ on the second site. On the second site, we also introduce a Dirac fermion $\Psi_{14}$ in $\bf 14$ (traceless symmetric) representation of the gauge symmetry $SO(5)_2$ as the heavy composite modes, with which the
$\Delta , \eta$ fermions will mix. In order to interact with $\Psi_{14}$, the $\Delta$ should also be embedded in the $\bf 14$ representation of the $SO(5)_1$ global symmetry:
\bea
 \Psi_\Delta = \frac{1}{2\sqrt{2}} \left(
   \begin{array}{ccc}
      -\sqrt{2} \Delta^0\mathds{1}_2  &  \Delta^{+-}  & \mathbf{0} \\
  (\Delta^{+-})^T  &    \sqrt{2} \Delta^0\mathds{1}_2  & \mathbf{0} \\
   \mathbf{0}& \mathbf{0} &0
  \end{array} \right),
\eea
where
\bea
  \Delta^{+-}=\left(
                \begin{array}{cc}
                  -\Delta^+-\Delta^- & i(\Delta^--\Delta^+) \\
                  i(\Delta^--\Delta^+) & \Delta^++\Delta^- \\
                \end{array}
              \right).
\eea

\begin{figure}
  \centering
  \includegraphics[width=7cm]{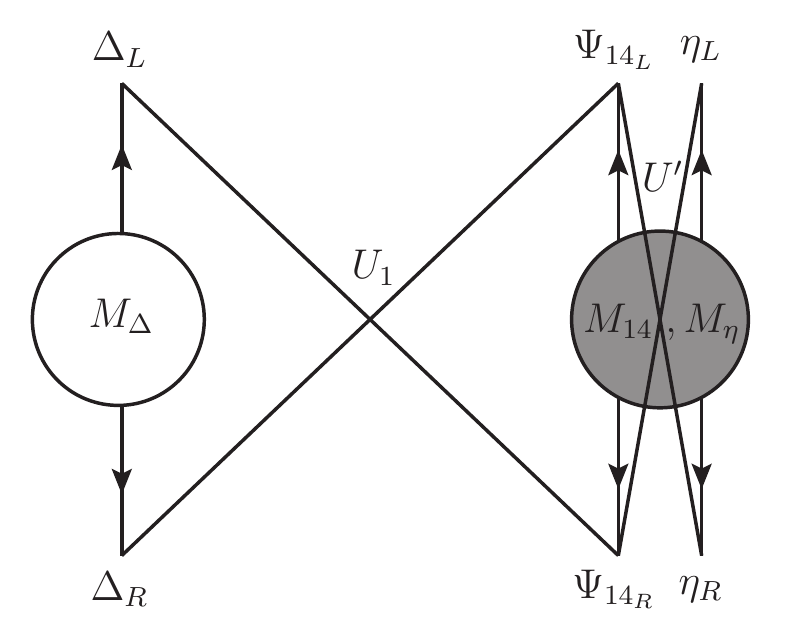}
  \caption{The triplet/singlet sector in two sites model.}\label{fig:2site_delta}
\end{figure}
The setup for this 2-site model is illustrated  in Fig~.\ref{fig:2site_delta}. The most general Lagrangian for the fermion fields invariant under the global $SO(5)_1 \times SO(5)_2$ is given by
\bea
&&  \mathcal{L}_{\Delta\eta}=\mbox{Tr}[\bar{\Delta}(i\slashed D-M_\Delta)\Delta ] + \bar{\eta}( i \slashed D -M_\eta)\eta  \nonumber \\
&&  + \mbox{Tr}[\bar{\Psi}_{14}( i \slashed D -M_{14})\Psi_{14}]
  -\Big( \lambda_{\Delta_L} \mbox{Tr}[\bar{\Psi}_{\Delta_L}U_1\Psi_{14_R}U_1^T]  \nonumber \\
  &&+\lambda_{\eta_R} \mathcal{H}'^\dagger  \bar{\Psi}_{14_L} \mathcal{H}' \eta_R
  +\lambda_{\Delta_R} \mbox{Tr}[\bar{\Psi}_{\Delta_R}U_1\Psi_{14_L}U_1^T]\nonumber \\
  &&+ \lambda_{\eta_L} \mathcal{H}'^\dagger  \bar{\Psi}_{14_R} \mathcal{H}' \eta_L +h.c. \Big).
\eea
The composite sector has an emlarged $SO(5)_{2L}\times SO(5)_{2R}$ global symmetry is broken by the fermion mass term and leaving behind the maximal symmetry $SO(5)_{2V}$. This maximal symmetry always allow some phase redefinitions on $\Psi_{14}$ to shift the pNGB fields. It actually preserves a shift symmetry in the triplet sector($\Psi_{14}\rightarrow U_1^T\Psi_{14}U_1$) and separately in the singlet sector($\Psi_{14}\rightarrow U'\Psi_{14}U'^T$). These shift symmetries are collectively broken by the triplet and singlet sector. After integrating out the $\Psi_{14}$, the only Higgs dependent term in the effective Lagrangian will be the triplet and singlet mixing term. Thus the Higgs potential must be proportional to $\sim (\lambda_\Delta  \lambda_\eta )^2$ and is only logarithmically divergent according to power counting.

Upon integrating out the massive $\Psi_{14}$, the effective Lagrangian for $\Delta$ and $\eta$  invariant under the global symmetry $SO(5)_1$ in momentum space is parametrized as
\begin{align} \label{eq:Ldelta_eff}
  \mathcal{L}_{\text{eff}}  &=  \Pi^0_{\Delta_L}
  \text{Tr}[\bar{\Psi}_{\Delta_L}\slashed{p}\Psi_{\Delta_L}]+\Pi^0_{\Delta_R}\text{Tr}[\bar{\Psi}_{\Delta_R}\slashed{p}\Psi_{\Delta_R}]  \nonumber\\
   & +\Pi^0_{\eta_L}  \bar{\eta}_L\slashed{p}\eta_L +\Pi^0_{\eta_R}\bar{\eta}_R\slashed{p} \eta_R\nonumber\\
   & - \Big(M^\Delta_0\text{Tr}[\bar{\Psi}_{\Delta_L}\Psi_{\Delta_R}]+M^\eta_0 \bar{\eta}_L\eta_R+h.c.\Big)\nonumber\\
    &+ \Big( \Pi^1_L \mathcal{H}^{\dagger} \bar{\Psi}_{\Delta_L} \mathcal{H} \slashed{p} \eta_L+\Pi^1_R \mathcal{H}^{\dagger} \bar{\Psi}_{\Delta_R} \mathcal{H} \slashed{p} \eta_R\nonumber\\
    &+M_1^{\Delta\eta}\mathcal{H}^{\dagger} \bar{\Psi}_{\Delta_L} \mathcal{H} \eta_R+M_2^{\Delta\eta}\mathcal{H}^{\dagger} \bar{\Psi}_{\Delta_R} \mathcal{H} \eta_L+h.c. \Big),
\end{align}
where the full expressions of form factors  are presented in Appendix~\ref{App:form_factor}.
We can see that only the mixing terms between triplet and singlet depend on the Higgs due to maximal symmetry. If  we express this effective Lagrangian in terms of the Higgs field, we find that all mixing terms are proportional to $s_h ^2$. Further integrating out the triplet $\Delta$ we obtain the effective Lagrangian for the singlet:
\begin{align}
  \mathcal{L}_{\text{eff}}^\eta&=(\Pi_{\eta_L}^0+\Pi_{\eta_L}^1s_h^4)\bar{\eta}_L\slashed{p}\eta_L+(\Pi_{\eta_R}^0
  +\Pi_{\eta_R}^1s_h^4)\bar{\eta}_R\slashed{p}\eta_R\nonumber\\
  &-(M^\eta_0+M^\eta_1s_h^4)(\bar{\eta}_L\eta_R + h.c.),
\end{align}
where the expressions of $\Pi_{\eta_L}^1$, $\Pi_{\eta_R}^1$ and $M^\eta_1$ are also presented in Appendix~\ref{App:form_factor}. The final Higgs potential from the above Lagrangian is
\begin{align}
  V_f &= -2 \int \frac{d^ 4 p_E }{(2\pi)^4} \log\Big[p_E^2 (\Pi_{\eta_L}^0+\Pi_{\eta_L}^1s_h^4)(\Pi_{\eta_R}^0+\Pi_{\eta_R}^1s_h^4)\nonumber\\
   &+ (M^\eta_0+M^\eta_1s_h^4)^2\Big].
\end{align}
For $s_h \ll 1$, we can expand the above Higgs potential and the leading term is
\begin{equation}
  V_f \approx \beta_\Delta  s_h^4,
\end{equation}
with
\begin{align}
 \beta_\Delta &= \frac{-2}{(4\pi)^2} \int_0^{\Lambda^2}dp_E^2\frac{p_E^2}{(M_0^\eta)^2+p_E^2\Pi^0_{\eta_L}\Pi^0_{\eta_R}}\times\nonumber\\
 &\Big(2M^\eta_0M^\eta_1+p_E^2(\Pi^0_{\eta_L}\Pi^1_{\eta_R}+\Pi^0_{\eta_R}\Pi^1_{\eta_L})\Big),
\end{align}
where $\Lambda \approx 4\pi f$ is strong dynamics confine scale.

Note that in the most general case, in addition to the kinetic mixing terms giving rise to a positive shift in the quartic we also obtain momentum independent left-right mixing  terms  in the effective Lagrangian (\ref{eq:Ldelta_eff}). These will generate a negative contribution to the Higgs quartic coupling as we discussed in Sec.\ref{Sec:brief_ideas}. In total there are four independent triplet-singlet mixing terms in the effective Lagrangian and any two of them can be contracted in a loop to give  contributions to the Higgs quartic coupling. The sum of  these contributions can be both positive and negative: the sign depends on the actual choices of the parameters in the model.  We uniformly scanned the whole parameter space and found that the region  corresponding to a positive $\beta_\Delta$ is large, which implies that in this model we can naturally obtain a positive shift Higgs for the quartic coupling. Details of this scan along the discussion of the remaining fine tuning will be presented in Sec.\ref{Sec:Fine_tune}, where we will show that $\beta_{\Delta}$ can be big enough to produce the observed Higgs mass with a small amount of tuning. The additional contributions to the Higgs potential from the top and gauge sectors are identical to those in models with minimal implementation of maximal symmetry, and are reviewed in App.~\ref{App:gauge} and~\ref{App:top}.

\section{Twin Higgs}\label{Sec:twin_Higgs}

Our mechanism of inducing a positive quartic is particularly interesting in the context of Twin Higgs Models~\cite{Chacko:2005pe,TH2,Craig,Geller:2014kta,Low:2015nqa,Barbieri:2015lqa}. Twin Higgs models (THM) solve the hierarchy problem by introducing a $Z_2$ parity $s_h\leftrightarrow c_h$ between the top and the twin top, where the twin top is $SU(3)_c$ color neutral. As a consequence of this $Z_2$ parity (which has been called Trigonometric Parity (TP) in\cite{Csaki:2017jby} and originates from the geometry of symmetric coset spaces) the color neutral twin top will cancel the quadratic divergences of the ordinary top, thereby realizing neutral naturalness. As a consequence ordinary (colored) top partners are expected to be very heavy and can easily evade bounds from direct searches.
 In order to achieve realistic EWSB, the $Z_2$ parity in Higgs potential must be broken (otherwise the Higgs VEV will be at $s_h^2\approx 0.5$). The usual approach to breaking this $Z_2$ is introducing an additional Higgs quadratic term, such as a $Z_2$-breaking gauge contribution\cite{Csaki:2017jby}, which partially cancels the quadratic term from the $Z_2$ preserving sector to ensure a small $s_h$. This cancellation is also the origin of the main tuning in THMs.

In this section we propose a novel way to break the $Z_2$ parity and achieve realistic EWSB in THMs, which will completely eliminate any leftover tuning in these models. We introduce our mechanism of generating a Higgs quartic coupling into the THMs and use this extra Higgs quartic term (rather than a usual quadratic term) for the source of the $Z_2$ breaking which will not introduce any tuning.

The minimal coset space that preserves the TP in the gauge sector is $SO(8)/SO(7)$. The EW gauge symmetry $SU(2)_L\times U(1)_Y$ and the twin EW gauge symmetry $SU(2)'_L\times U(1)'_Y$ are separately embedded in the $SO(4)_1$ and $SO(4)_2$ subgroups which act on the first four and last four indices of $SO(8)$ respectively. The 2-site implementation of this $SO(8)/SO(7)$ twin Higgs is similar to the $SO(5)/SO(4)$ case shown in Appendix.\ref{App:gauge}. The divergences from the one loop EW gauge contributions to the Higgs potential are cancelled by their twin partners due to the $Z_2$ symmetry, and the leading order Higgs potential will be $\mathcal{O}(g^4)$  and of the form
\begin{equation}
  V_g(h)\approx -\beta_g(s_h^4+c_h^4),
\end{equation}
where $\beta_g$ can be parameterized as
\begin{equation}
 \beta_g=c_g\frac{g^4f^4}{(4\pi)^2}\log\frac{m_\rho^2}{m_W^2},
\end{equation}
with $c_g$ a numerical constant.

In order to preserve the $Z_2$ TP in the fermion sector, color neutral twin tops $\tilde{q}_L, \tilde{t}_R$ are introduced which transform under twin color $SU(3)'_c$. More details of the construction of $SO(8)/SO(7)$ THM are shown in Appendix.\ref{App:twin_Higgs}. The leading contribution to the Higgs potential from the top-twin top sector will be at $\mathcal{O}(y_t^4)$,
\begin{equation}
  V_t(h)\approx \beta_f(s_h^4+c_h^4),
\end{equation}
with
\begin{equation}\label{eq:betaf_count}
  \beta_f=c_f \frac{N_cy_t^4f^4}{(4\pi)^2}\log\frac{M_f^2}{m_t^2}.
\end{equation}
The quadratic divergences will be canceled by the twin partners both in the gauge and the fermion sectors hence $\beta_{g,f}$ will only be logarithmically sensitive to the mass of the  vector mesons and colored top partners. Hence the Higgs can be light even for heavy colored top partners
with only a mild logarithmic tuning. However to achieve a realistic EWSB minimum away from the $Z_2$ symmetric point $s_h=c_h = \frac{1}{\sqrt{2}}$ one needs to explicitly break the $Z_2$ symmetry, which is usually done by introducing an explicit breaking in the gauge sector,  leading
to gauge contributions  $\mathcal{O}(g^2)$, $V_g\sim f^2g^2m_\rho^2$. This is much bigger than the $Z_2$ preserving term and also sensitive to the gauge partner mass of $m_\rho$. Usually this is also the leading source of the tuning in TH models. Our mechanism of generating the additional quartic will be able to significantly reduce this tuning. In our approach we leave the gauge sector $Z_2$ invariant and the only source of $Z_2$ breaking will be in the fermion sector responsible for the generation of the quartic.  As before, the $SU(2)_L$ triplet $\Delta$ will be embedded in the traceless symmetric representation, which in the case of $SO(8)$ is a $\mathbf{35}$. Just like before the $\Delta$ will be added at the first $SO(8)$ site while the singlet $\eta$ at the second site. The explicit embedding of $\Delta$ in $\mathbf{35}$ is
\begin{equation}\label{eq:embed_35}
  \Psi_\Delta=\frac{1}{2\sqrt{2}} \left(
   \begin{array}{ccc}
      -\sqrt{2} \Delta^0\mathds{1}_2  &  \Delta^{+-}  & \mathbf{0} \\
  (\Delta^{+-})^T  &    \sqrt{2} \Delta^0\mathds{1}_2  & \mathbf{0} \\
   \mathbf{0}& \mathbf{0} &0_{4\times 4}
  \end{array} \right).
\end{equation}
It is the fact that we only introduce a single $\Delta$ triple (and no twin $\Delta$ that could be in the lower right 4 by 4  corner of the above matrix) that is the source of $Z_2$ breaking, which will eventually lead to a shift in the quartic $s_h^4$ term (but no analogous $c_h^4$ term). To complete the construction we should also introduce a Dirac fermion $\Psi_{35}$ in the $\mathbf{35}$ representation of $SO(8)_2$ which will mix  with $\Delta$ and $\eta$ as explained in the previous sections. The rest of the construction is completely analogous to the $SO(5)/SO(4)$ case. After integrating out all the fermions we will get an adjustable Higgs quartic term and the form of the Higgs potential is
\begin{equation}
  V(h)=(\beta_f-\beta_g)(s_h^4+c_h^4)+\beta_\Delta s_h^4.
\end{equation}
Clearly the extra quartic term $\beta_\Delta$ breaks the $Z_2$ symmetry of the Higgs potential, and will give rise to a lrealistic minimum with no tuning at all as we will explicitly demonstrate in Sec.~\ref{Sec:Fine_tune}.

\section{Higgs potential and fine tuning}\label{Sec:Fine_tune}

In this section, we will discuss the properties of the Higgs potential in the two models with extra Higgs quartic terms presented above. We show that the fine tuning needed to obtain realistic EWSB is very significantly reduced due to our mechanism for generating the Higgs quartic.

\subsection{Model with minimal maximal symmetry}

In composite Higgs models the Higgs potential can be parametrized as
\bea
V(h) =- \gamma s_h^2 +\beta s_h^4,
\eea
where it is assumed that for realistic EWSB VEV $s_h \ll 1$ hence higher powers of $s_h$ are neglected. The coefficients $\gamma =\gamma_f -\gamma_g$, $\beta =\beta_f +\beta_\Delta$ include the fermion (f) and gauge (g) sector contributions, and we already added the extra quartic contribution from our mechanism $\beta_\Delta$. The overall $\beta$ has to be positive for a realistic model so the pNGB Higgs will acquire  a VEV if $\gamma > 0$ with a minimum at
\bea
s_h^2= \frac{\gamma}{2\beta}\equiv\xi,
\eea
where $\xi$ is a parameter measuring the separation between $f$ and EWSB scale $v$. The Higgs mass in this vacuum is
\bea \label{eq:Higgs}
m_h^2 =\frac{8\beta \xi (1-\xi)}{f^2}.
\eea
Using $v^2/f^2 \sim \xi $, we see that the Higgs mass depends only on $\beta/f^4$. The value of $\beta$ reproducing $m_h=125$ GeV is
\bea
\frac{\beta}{f^4} \approx 0.036\ .
\label{betaval}
\eea

We can now calculate the tuning in the model with minimal maximal symmetry and consider the effect of the additional quartic. The expressions of $\gamma_f, \beta_f$ in this model are (see  Eq.~(\ref{eq:gabe_fermion}) )
\begin{equation}\label{eq:Appr_gabef}
 \gamma_f \simeq\frac{2N_c y_t ^2 M_f ^2 f^2}{(4\pi)^2} ,\;\beta_f \simeq \frac{N_c y_t^4f^4}{(4\pi)^2 } \ln\frac{M_f^2}{m_t^2},
\end{equation}
where $M_f$ is a typical top partner mass. Without the additional contribution $\beta_\Delta$ to the quartic the model needs to have extremely heavy top partners because $\beta_f$ has only a logarithmic dependence on $M_f$.  If we fix $\xi$ to 0.1 and $m_t\in[140,170]$ GeV, a rough bound on $M_f$ is about $M_f\gtrsim 11$ TeV to obtain a sufficiently heavy Higgs.   Since $\gamma_f$ is much bigger than $\beta_f$, $\gamma_f$ must first be tuned to be of order $\beta_f$ and then further tuned to $\xi\beta$ which results in double tuning that can be quantified as
\begin{equation}
  \Delta=\left|\frac{\partial\ln\xi}{\partial\ln M_f}\right|\approx \frac{\gamma_f}{\xi\beta_f}=\frac{1}{\xi}\frac{2M_f^2}{f^2y_t^2\ln\frac{M_f^2}{m_t^2}}.
\end{equation}
The strong bound on $M_f$ eventually results in a large tuning about $\Delta\gtrsim 95/\xi$, numerically about 0.1 percent or worse.

\begin{figure}[tp]
  \centering
  \includegraphics[width=7.1cm]{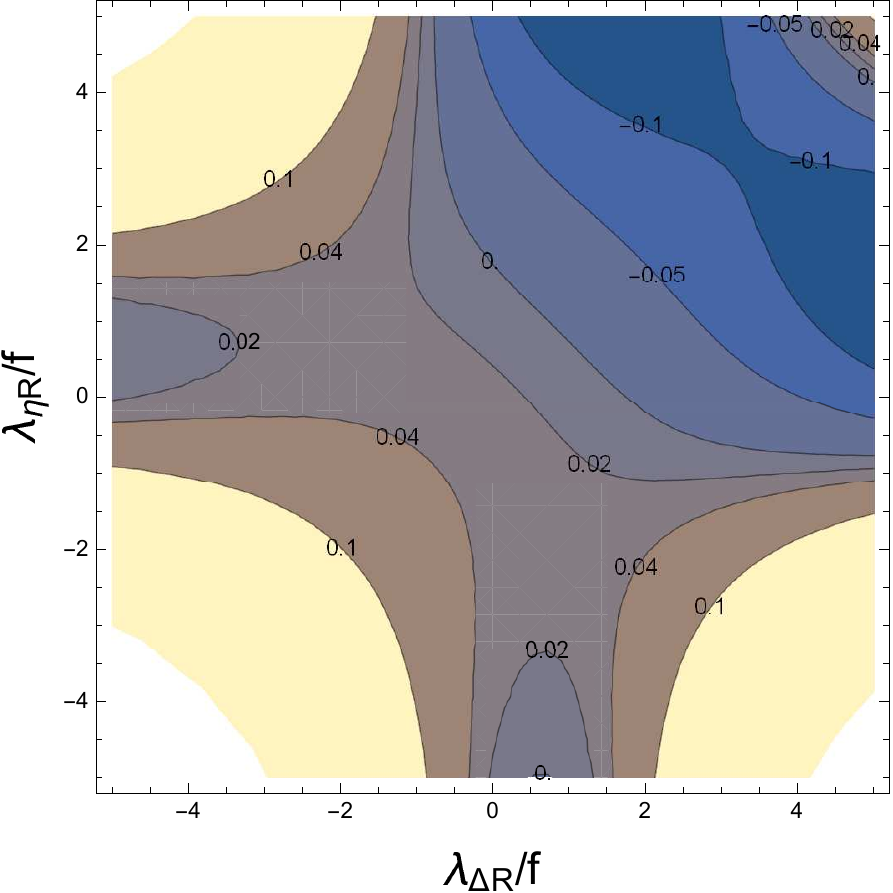}\\
  \caption{Contour plot for the Higgs quartic coupling $\beta_\Delta/f^4$ as a function of the right-handed couplings $\lambda_{\Delta_R}/f$ and $\lambda_{\eta_R}/f$ with the masses of the triplet, the singlet and the $\Psi_{14}$ multiplets  fixed at 4 TeV, $M_\Delta=M_\eta=M_{14}=4$ TeV, and the left-handed couplings fixed at   $\lambda_{\Delta_L}=2f$, $\lambda_{\eta_L}=2f$.}\label{fig:beta_delta}
\end{figure}

The addition of the extra quartic however improves the situation greatly, to the point that this model will need the smallest amount of tuning of composite Higgs models with heavy top partners. Before quantifying the tuning we first illustrate that our mechanism can indeed produce a sufficiently large positive Higgs quartic coupling. To demonstrate that we can easily achieve
the value $\beta$ from (\ref{betaval})  we  show a contour plot of $\beta_\Delta/f^4$ in part of the right-handed coupling parameter space in Fig.\ref{fig:beta_delta}, where we fixed the bare masses of $\Psi_\Delta$, $\Psi_\eta$ and $\Psi_{14}$ at $4$ TeV and also fixed the left-handed coupling to an appropriate value, $(\lambda_{\Delta_L},\lambda_{\eta_L})=(2f,2f)$. We see that for a sizeable fraction of the parameter space $\beta_\Delta$ is sufficiently large to produce the observed Higgs mass.

\begin{figure}
\begin{center}
\includegraphics[width=0.8\columnwidth]{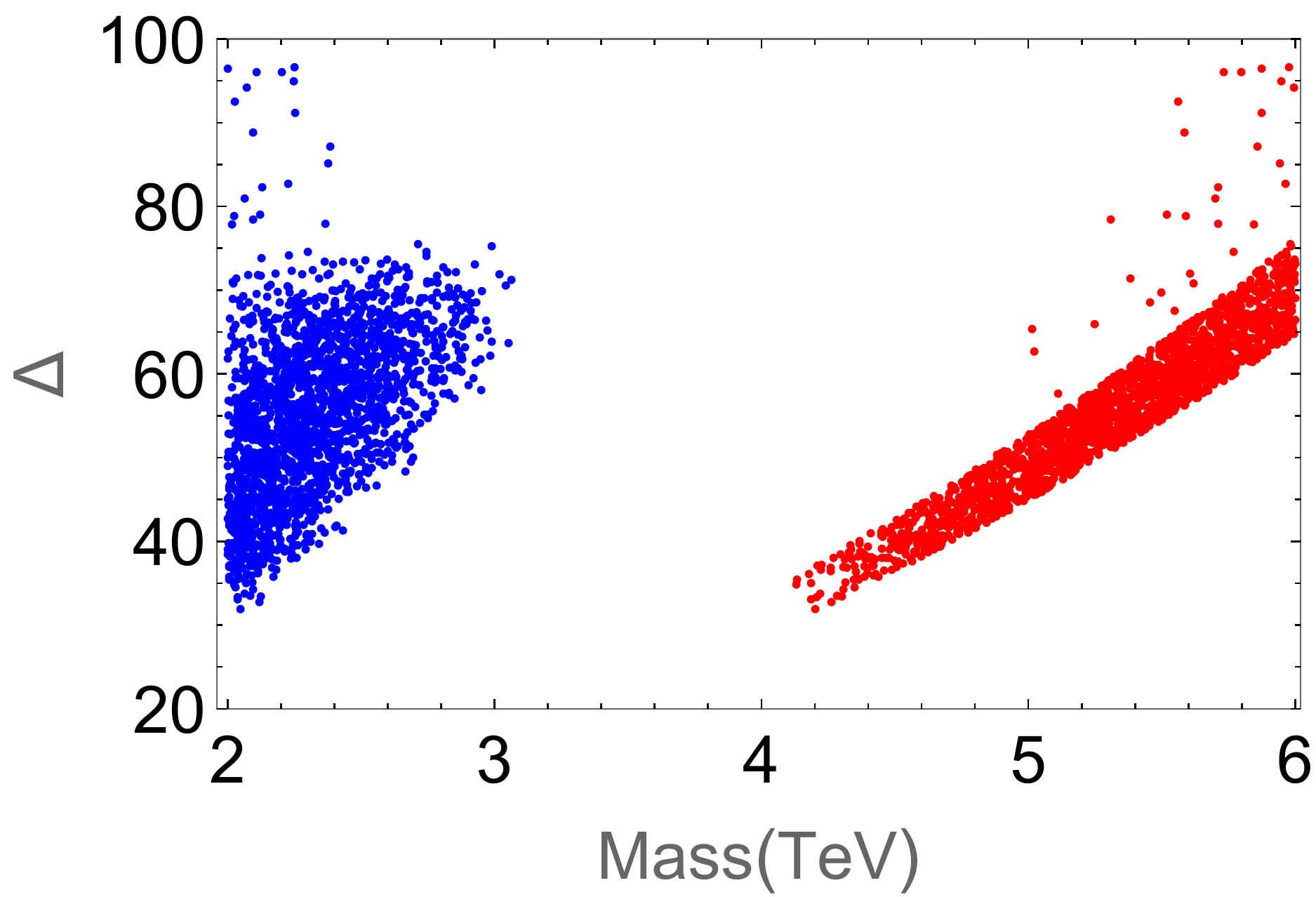}
\end{center}
\caption{Scatter plot of the tuning $\Delta$ in the model with the extra quartic and minimal maximal symmetry
as function of the top partner mass (Blue) and gauge boson mass (Red) for $\xi =0.1$. The Higgs mass is fixed at 125 GeV and the top mass range is $m_t \in [140, 170]$ GeV. We require that  the lightest top partner mass $M >2$ TeV and gauge boson mass $m_\rho >2.5$ TeV.}
\label{fig:tuning}
\end{figure}

With the addition of this adjustable Higgs quartic coupling, the tuning becomes
\begin{equation}
  \Delta' \approx \frac{\gamma_f}{\xi(\beta_f+\beta_\Delta)}=\Delta\frac{M^{\prime 2}_f}{M_f^2},
\end{equation}
where we use $M_f^{\prime}$ ($M_f$) to denote the typical top partner mass for achieving right Higgs mass after (before) $\beta_\Delta$ is added and their relation is $M_f =M_f^{\prime} (M_f^\prime/m_t)^{\frac{\beta_\Delta}{\beta_f}}$. Notice that $\beta$ is actually fixed by the Higgs mass so the enhancement of $\beta_\Delta$ will reduce $\beta_f$ which will result in a great reduction of the top partner mass. Hence the original double tuning will be strongly suppressed by $M_f^{\prime}/M_f$. In terms of the physical parameters the tuning will be
\begin{equation}\label{eq:tune_deltaprime}
  \Delta'=\frac{1}{\xi}\frac{N_c(1-\xi)m_t^2g_f^{\prime 2}}{\pi^2m_h^2},
\end{equation}
where $g'_f\equiv M'_f/f$. If we require  $M_f^{\prime}$ to be at least 2 TeV, the tuning will be $\Delta'\gtrsim 2.3/\xi$, which for $\xi =0.1$ corresponds to a roughly 5\% tuning. This rough estimate can be verified by the numerical scan in Fig.\ref{fig:tuning}, where we show the tuning as function of the top partner and gauge partner masses for $\xi=0.1$ with Higgs mass fixed $m_h =125$ GeV and $m_t \in [140,170] $ GeV.

As a comparison we present the estimate for tuning in the more standard holographic composite Higgs models or their deconstructed versions. Generically these models have a double tuning because $\gamma_f$ appears at order $\mathcal{O}(y_t)$ and $\beta_f$ at order $\mathcal{O}(y_t^2)$. In addition these models also require a non-generic top partners spectrum in which some of the top partners are anomalously light with mass $M_p$
below  the typical top partner mass scale $M_f$. Such states are needed to ensure that the  Higgs is sufficiently light~\cite{Panico:2012uw}. In fact the typical value of the lightest top partner mass is generically below 1 TeV, with an upper bound of around 1.5 TeV for $\xi$ fixed at 0.1 in these models~\cite{Panico:2012uw,Matsedonskyi:2012ym}. For example in \cite{Matsedonskyi:2012ym}  a single light top partner of  600 GeV was assumed in a three-site model. Top partners this light are excluded, since the most recent bounds on generic top partners is of order 1.5 TeV. The only way to make these models viable is by lowering the value of  $\xi$ and thereby raising the mass scale of all partners, at the price of increasing the tuning.  The tuning in the 5d holographic models is~\cite{Panico:2012uw} (assuming the sub-TeV top partners)
\begin{equation}
  \Delta^{\mathbf{5}+\mathbf{5}}_{\xi}=\frac{1}{\xi}\sqrt{\frac{N_c}{2\pi^2}}\frac{g_f^2v}{m_h}\ ,
\end{equation}
where $g_f=M_f/f$. The additional tuning due to the lowering of $\xi$ is
\begin{equation} \label{eq:tuning_scale}
  \Delta^{\mathbf{5}+\mathbf{5}}_{\xi^\prime}=\Delta^{\mathbf{5}+\mathbf{5}}_{\xi}\frac{M_p^{\prime 2}}{M_p^2},
\end{equation}
where $M_p$ is the mass scale of the lightest sub-TeV top partner in the original model which is enhanced to $M'_p$ after $\xi$ is lowered.
Numerically we find  $\Delta^{\mathbf{5}+\mathbf{5}}_{\xi^\prime}\approx 2\Delta' M_p^{\prime 2}/M_p^2$ where $\Delta'$ is the tuning in our model with minimal maximal symmetry and the additional quartic (see Eq.\ref{eq:tune_deltaprime} and take $g_f\approx g'_f$). If we set $M'_p$ to 2 TeV and choose $M_p$ to 1 TeV for $\xi =0.1$, we find $\Delta^{\mathbf{5}+\mathbf{5}}_{\xi^\prime}\approx 8\Delta'$. Hence our model has at least about 8 times less tuning than the holographic/deconstructed composite Higgs models with comparable parameters. Note that increase of the experimental bound on $M'_p$ will increase the ratio between the tuning in holographic CHM's vs. our model presented here.

Finally we discuss the tuning for the case of ordinary maximal symmetry~\cite{Csaki:2017cep}. For this model both $\gamma_f$ and $\beta_f$ are at the order $\mathcal{O}(y_t^2)$ and approximately equal, $\gamma_f\approx\beta_f$. Hence this model doesn't have double tuning and its tuning is minimal about $1/\xi$,
\begin{equation}
  \Delta_\xi\simeq\frac{\gamma_f}{\xi\beta_f}= \frac{1}{\xi}.
\end{equation}
Note however that since $\beta_f$ is ${\cal O}(y_t^2)$, the Higgs mass is sensitive to the top partner masses. Hence a light top partner is  required in order to get a light Higgs. The typical value of the lightest top partner mass is below 1 TeV for $\xi$ fixed to 0.1, which is already excluded by the most recent LHC searches. To raise the top partner mass one has no choice but to reduce $\xi$ to $\xi^\prime$ with $\xi^\prime/\xi = M_{p}^{2}/M_p^{\prime 2}$ as in the previous case. Hence, similar to Eq.~(\ref{eq:tuning_scale}), the final tuning in the ordinary maximally symmetric model will be
\begin{equation}
  \Delta_{\xi^\prime}\simeq \frac{1}{\xi^\prime}  = \Delta_\xi \frac{M_p^{\prime2}}{M_{p}^2}\ .
\end{equation}

If we add our Higgs quartic mechanism to this model, the tuning will be reduced to
\begin{equation}
  \Delta_{\xi} \simeq\frac{1}{\xi}\frac{\gamma_f}{\beta_f+\beta_\Delta}=\frac{1}{\xi}\frac{M_{F}^2}{M_{p}^2},
\end{equation}
where $M_F$ is the lightest top partner mass after $\beta_\Delta$ is added. Note that $M_F$ is smaller than $M_{p}$ because the addition of $\beta_\Delta$ will always reduce the top partner masses. Hence we will now reduce the $\xi$ of the model even further, by a factor of $M_F^2/M_{p^\prime}^2$. The final resulting tuning will be again
\begin{equation}
  \Delta'\simeq\frac{1}{\xi} \frac{M_{p^\prime}^2}{M_{p}^2}
\end{equation}
as for the ordinary maximally symmetric model with the difference that the model with the additional quartic has a smaller $\xi$ (for the same top partner spectrum). Hence it corresponds to weaker couplings and smaller corrections to the Higgs branching ratios. The reason why the quartic generation mechanism is not very effective for the case of ordinary maximal symmetry is that this model doesn't have heavy top partners.
  Generically the Higgs quartic mechanism can be used in any pNGB Higgs model with maximal symmetry to reduce the tuning. But the amount of reduction in the tuning depends on how heavy the top partners are:    The models with the heaviest top partners will enjoy the biggest gains in the tuning due to the addition of $\beta_\Delta$.

\subsection{Twin Higgs model}

Our mechanism for  generating a Higgs quartic has very beneficial effects in THM's. Before we explicitly estimate the effects on the tuning we would like to emphasize that the effect of the quartic on THM's is different from the cases considered before. THM's have a $Z_2$ symmetry which will soften the Higgs potential by eliminating the leading order contributions, which will also greatly reduce the tuning. However for achiving realistic EWSB the $Z_2$ has to be broken either in the gauge or the top sector, which will reintroduce the sensitivity to the partner masses and some of the tuning into the model. The beauty of our mechanism of generating the quartic is that this could also be the source of $Z_2$ breaking, allowing the top and gauge sectors to remain exactly $Z_2$ invariant and without introducing any tuning. In essence the Higgs quartic mechanism will ensure that the $Z_2$ symmetry of the gauge and top sectors has the maximal effect on softening the Higgs potential.

To examine the effect on the tuning in detail we assume that the origin of the $Z_2$ breaking is from our mechanism of generating the extra quartic. Then the Higgs potential can be parametrized as
\bea
\gamma =2(\beta_f -\beta_g), \quad \beta= 2(\beta_f -\beta_g)+\beta_\Delta.
\eea

In this case, the Higgs VEV will be at
\bea
\xi =\frac{1}{2(1+x)}, \; \;  x =\frac{\beta_\Delta}{2( \beta_f -\beta_g)}.
\eea

For a TH without $Z_2$ breaking the Higgs VEV would be at $\xi =0.5$. The main effect of the $Z_2$ symmetry is to soften the Higgs potential by eliminating the leading order contributions (and hence the main dependence on the top/gauge partner masses). Once the $Z_2$ is broken to allow for realistic EWSB, the dependence on the top/gauge partner will be reintroduced, enhancing $\gamma_f$ (or $\gamma_g$). To obtain a small $\xi$ we will then need some cancellation between $\gamma_f$ and $\gamma_g$ leading to the main source of tuning in this model.

 This is in sharp contrast to the situation in our new model where the $Z_2$ parity is broken by the additional Higgs quartic term, which also results in the enhancement of $\beta$. In this model
a small $\xi$ can be achieved without breaking the $Z_2$ in the top and gauge sectors. Both $\gamma_f$ and $\gamma_g$ are vanishing at the leading order due to the exact $Z_2$ symmetry in the top and gauge sectors. Hence both $\gamma_f$ and $\gamma_g$ remain small
 and insensitive to the partner masses, unlike in ordinary twin Higgs models where the $Z_2$ breaking raises $\gamma_{f,g}$ back the leading order. For example, with $Z_2$ breaking in the gauge sector in the SO(6)/SO(5) TH model of~\cite{Csaki:2017jby} $\gamma_g$ can be estimated as
\begin{equation}
 \gamma_g^{TH}\simeq \frac{3f^2(3g^2+g^{\prime 2}-2g_1^2)m_\rho^2\ln2}{64\pi^2}\ ,
\end{equation}
where $g_1$ is a gauge coupling of an additional U(1)$_\eta$. In our model with the additional quartic we find using~(\ref{eq:gauge_potential})
\begin{equation}
\gamma'_g=2\beta_g\simeq \frac{9f^4g^4}{512\pi^2}\ln\frac{m_\rho^2}{m_W^2}\ .
\label{eq:THgamma}
\end{equation}
We can see that this model is in fact not tuned at all, but provides an example of a viable fully natural electroweak symmetry breaking. There could potentially be two sources of tuning - in obtaining the desired values of $\beta_\Delta$ and $\gamma$. Evaluating the tuning for the quartic we obtain with $\lambda_\Delta \equiv \beta_\Delta/ f^4$
\bea
\Delta_\lambda =\frac{\beta_\Delta}{\beta}=1-2\xi.
\eea
Numerically we find $\Delta_\lambda \simeq 0.7$ for $\xi =0.15$ and with the lightest top partner mass above 2 TeV. Hence there is no tuning at all. The other potential source of tuning would be from the dependence of $\gamma$ on the resonance masses, but as we see in (\ref{eq:THgamma}) that there is only a mild logarithmic dependence on $m_\rho^2$ hence this will not be tuned either, yielding a fully natural Higgs potential.

\begin{figure}
\begin{center}
\includegraphics[width=0.49\columnwidth]{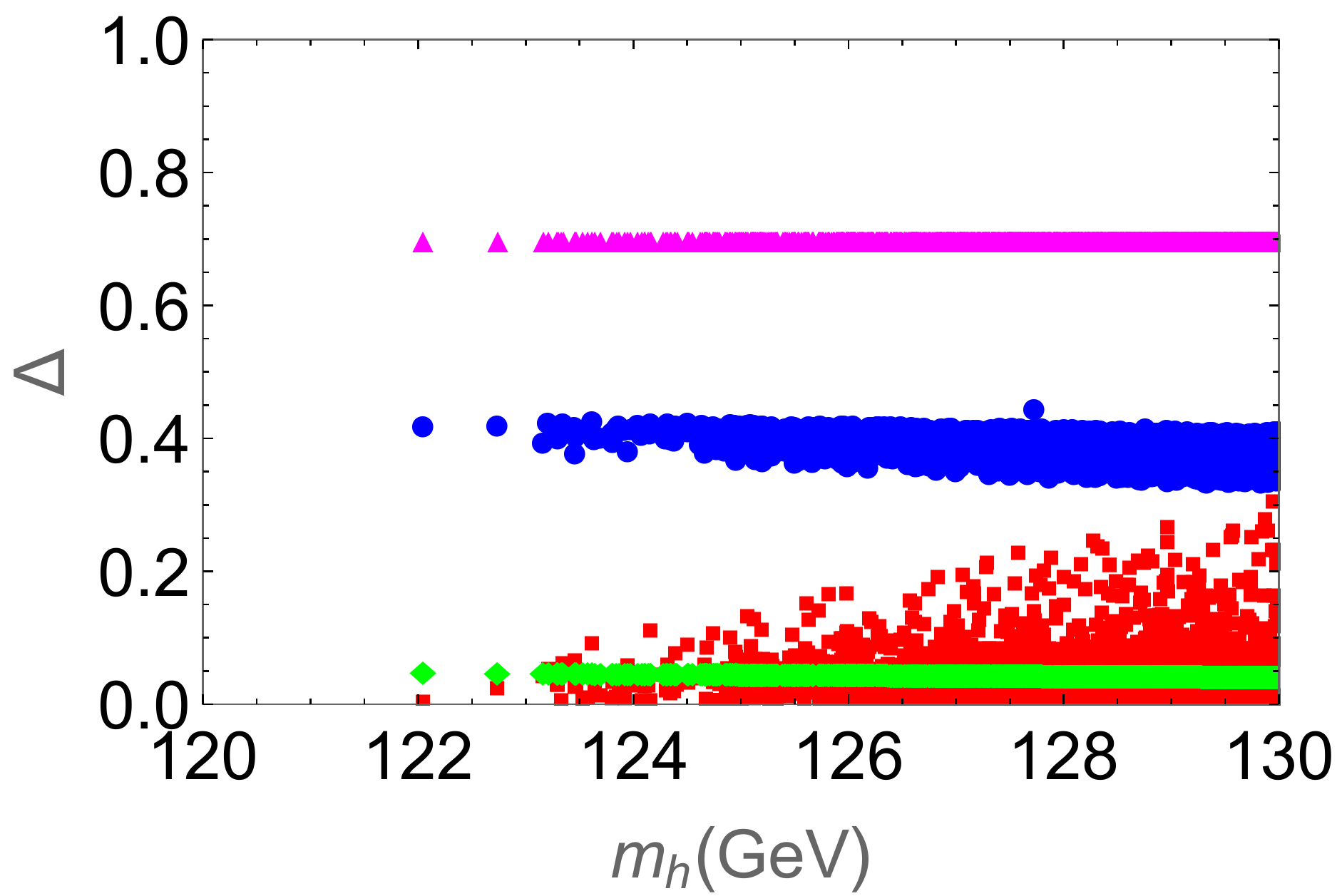}
\includegraphics[width=0.49\columnwidth]{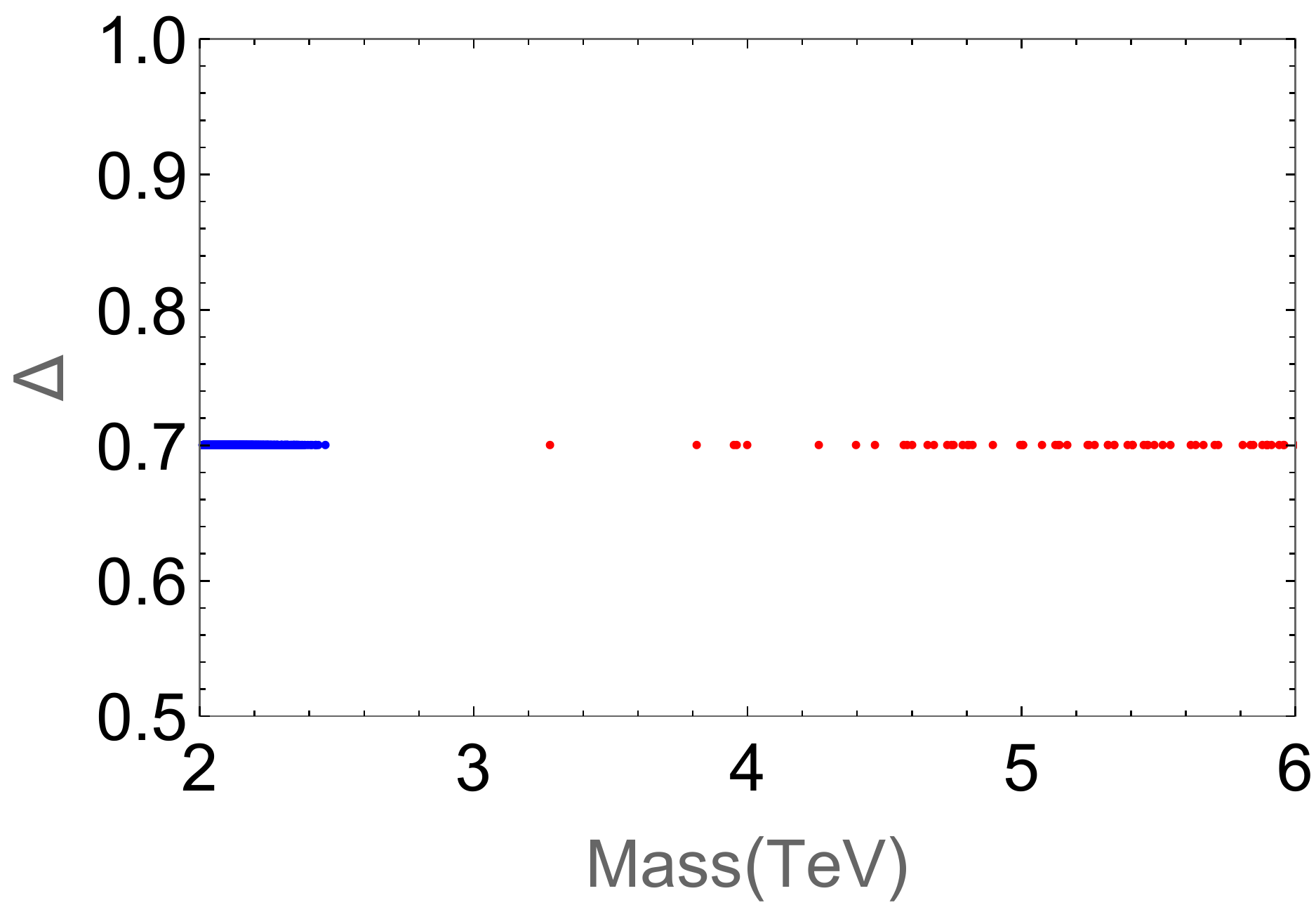}
\end{center}
\caption{Scatter plot of the tuning in THMs for $\xi =0.15$ with the Higgs mass fixed at 125 GeV and the top mass range is $m_t \in [140, 170]$ GeV. The left panel shows the tuning from different parameters $M$(blue), $\epsilon_R$(red), $m_\rho$(green) and $\beta_\Delta$(magenta) as a function of the  Higgs mass. The right panel is the tuning $\Delta$ as function of top partner mass (Blue) and gauge boson mass (Red). We restrict the top partner mass $M >2$ TeV and gauge boson mass $m_\rho >2.5$ TeV.}
\label{fig:tuning_twin}
\end{figure}

We would like to compare this situation numerically with the tuning in ordinary TH. The main tuning in ordinary TH comes from the sensitivity to $m_\rho$   and is a constant when the mass of $\rho$ meson is light (lower than 3 TeV),
\begin{equation}
  \Delta^{TH}=\frac{1-2\xi}{\xi}.
\end{equation}
So the tuning in ordinary TH is about 7 times bigger than in our model for $\xi =0.15$, numerically $\Delta^{TH}\sim 5$. Moreover, things will become slightly worse in ordinary TH when $m_\rho$ increases because the cancellation between contributions of the gauge sector and the twin gauge sector will become significant, which result in a notable increase in the tuning which will grow  linearly with $m_\rho^2$.
In Fig.~\ref{fig:tuning_twin}, we show the tuning as a function of the partner masses for $\xi=0.15$ with the Higgs mass fixed to $m_h =125$ GeV and $m_t \in [140,170] $ GeV.

\section{Conclusions\label{Sec:conclusion}}

An adjustable Higgs quartic self-coupling can play an important role in producing a natural Higgs potential and solving the little hierarchy problem. In this work we proposed a novel mechanism for producing such an adjustable Higgs quartic term.
It is based on the observation that a kinetic mixing between EW singlet and triplet fermions can result in a positive contribution to the Higgs quartic. This mechanism is very simple and can be implemented in any composite pNGB Higgs model. We presented an explicit realization in a two site MCHM with the minimal implementation of maximal symmetry as well as in a simple Twin Higgs model. In the first model, the Higgs quartic term from the gauge and top sectors is not big enough to produce the observed Higgs mass without enormously heavy top partners which then feed into the Higgs quadratic term. The additional Higgs quartic allows to easily reproduce the observed Higgs mass without ultra heavy top partners hence significantly reduces the tuning needed for successful EWSB. The role of the additional quartic is somewhat different in the Twin Higgs model. In these models it can be used as the (only) source for  $Z_2$ breaking while keeping both the  gauge and top sectors $Z_2$ invariant. As a result the Higgs potential will be largely insensitive to colored top and gauge partner masses. These  Twin Higgs models will have a fully natural EWSB sector with  the heavy colored partners outside of the LHC direct detection bounds.

\section*{Acknowledgements}

C.C. thanks the Technical University of Munich for a fruitful visit while working on this project supported by a research prize by the Humboldt Research Foundation. T.M. thanks the Cornell Particle Theory group for its hospitality while working on this project. C.C. is supported in part by the NSF grant PHY-1719877 as well as the BSF grant 2016153. J.S. is supported by the National Natural Science Foundation of China (NSFC) under grant No.11847612, No.11690022, No.11851302, No.11675243 and No.11761141011, and also sup- ported by the Strategic Priority Research Program of the Chinese Academy of Sciences under grant No.XDB21010200 and No.XDB23000000. T.M. is supported in part by project Y6Y2581B11 supported by 2016 National Postdoctoral Program for Innovative Talents.

\appendix
\section{Form Factors}
\label{App:form_factor}
In this section we present the explicit form factors appearing in the effective Lagrangian. The form factors in the triplet-singlet effective Lagrangian are
\begin{align}\label{eq:delta_formfac}
  \Pi^0_{\Delta_L} & =1+\frac{\lambda_{\Delta_L}^2}{M_{14}^2-p^2},\quad \Pi^0_{\Delta_R}  =1+\frac{\lambda_{\Delta_R}^2}{M_{14}^2-p^2}, \nonumber\\
  \Pi^0_{\eta_L} & =1+\frac{4\lambda_{\eta_L}^2}{5(M_{14}^2-p^2)},\quad \Pi^0_{\eta_R}  =1+\frac{4\lambda_{\eta_R}^2}{5(M_{14}^2-p^2)},\nonumber\\
  M^\Delta_0&= M_\Delta-\frac{M_{14}\lambda_{\Delta_L}\lambda_{\Delta_R}}{M_{14}^2-p^2},\: M^\eta_0= M_\eta-\frac{4M_{14}\lambda_{\eta_L}\lambda_{\eta_R}}{5(M_{14}^2-p^2)},\nonumber\\
  \Pi^1_L & =\frac{\lambda_{\Delta_L}\lambda_{\eta_L}}{M_{14}^2-p^2},\quad\Pi^1_R =\frac{\lambda_{\Delta_R}\lambda_{\eta_R}}{M_{14}^2-p^2},\nonumber\\
  M^{\Delta\eta}_1&=\frac{M_{14}\lambda_{\Delta_L}\lambda_{\eta_R}}{M_{14}^2-p^2},\quad M^{\Delta\eta}_2=\frac{M_{14}\lambda_{\Delta_R}\lambda_{\eta_L}}{M_{14}^2-p^2}.
\end{align}
The form factors in the singlet effective Lagrangian are defined by
\begin{align}
  \Pi^1_{\eta_L} & = \frac{\Pi^0_{\Delta_L}(M^{\Delta\eta}_2)^2+2M_0^\Delta\Pi^1_LM^{\Delta\eta}_2+p^2\Pi^0_{\Delta_R}(\Pi^1_L)^2}
  {4((M^\Delta_0)^2-p^2\Pi^0_{\Delta_L}\Pi^0_{\Delta_R})}, \nonumber\\
  \Pi^1_{\eta_R} & = \frac{\Pi^0_{\Delta_R}(M^{\Delta\eta}_1)^2+2M_0^\Delta\Pi^1_RM^{\Delta\eta}_1+p^2\Pi^0_{\Delta_L}(\Pi^1_R)^2}
  {4((M^\Delta_0)^2-p^2\Pi^0_{\Delta_R}\Pi^0_{\Delta_L})}, \nonumber\\
  M^\eta_1&=\frac{1}{4(p^2\Pi^0_{\Delta_L}\Pi^0_{\Delta_R}-(M^\Delta_0)^2)}\times\nonumber\\
  &\:\Big(p^2\Pi^1_L(\Pi^0_{\Delta_R}M^{\Delta\eta}_1+M^\Delta_0\Pi^1_R)\nonumber\\
  &+M^{\Delta\eta}_2(M^\Delta_0M^{\Delta\eta}_1+p^2\Pi^0_{\Delta_L}\Pi^1_R)\Big).
\end{align}

\section{Gauge sector}\label{App:gauge}
In this section, we present the details of the gauge sector in the 2-site $SO(5)/SO(4)$ model. The NGB fields  can be parameterized  as
\bea
  U_1=\text{Exp}(\frac{i\pi_1^aT^a}{f}),\quad U'=\text{Exp}(\frac{i\pi_2^{\hat{a}}T^{\hat{a}}}{f}),
\eea
where $T^a$ is the generator of $SO(5)$ and $T^{\hat{a}}$ is the one in $SO(5)/SO(4)$ with the normalization of $\text{Tr}[T^a T^b] =\delta^{ab}$.
For simplicity we have assumed that the decay constants of the two link fields are equal. The gauge interactions for the two link fields $U_1$ and $\mathcal{H}^\prime$ are given by
\bea\label{eq:gauge_Lag}
  \mathcal{L}_g&=\frac{f^2}{2}\text{Tr}[D_\mu U_1(D^\mu U_1)^\dag]+f^2(D_\mu^\prime \mathcal{H}')^\dag D^{\prime \mu}\mathcal{H}'\nonumber\\
  &-\frac{1}{4}\text{Tr}[ \rho_{\mu\nu}\rho^{\mu\nu}]-\frac{1}{4}W_{\mu\nu}^a W^{\mu\nu,a}-\frac{1}{4}B_{\mu\nu}B^{\mu\nu},
\eea
\noindent where $\rho_{\mu\nu} \equiv  \rho_{\mu\nu}^a T^a$ is the $SO(5)_2$ gauge boson, the covariant derivatives are $D_\mu U_1=\partial_\mu U_1-i(g A_\mu^aT_L^a+g' B_\mu T_R^3)U_1+ i g_\rho U_1\rho_\mu^aT^a$, $D_\mu^\prime \mathcal{H}'=\partial_\mu \mathcal{H}'-ig_\rho \rho_\mu^aT^a \mathcal{H}'$ and $T_L^a$ and $T_R^3$ are $SU(2)_L$ and $U(1)_Y$ generators embedded in the $SO(4)$ custodial symmetry. After EWSB, only the Higgs boson is remains uneaten. In unitary gauge, after integrating out the gauge resonances $\rho_\mu$, we get the effective Lagrangian of the gauge fields (at quadratic order) interacting with Higgs boson  as
\begin{align} \label{eq:Lg_eff}
 \mathcal{L}_g^{\text{eff}}&= \frac{P_t^{\mu\nu}}{2} \Big[\Pi_W W_\mu^a W_\nu^a+\Pi_1\frac{s_h^2}{4}(W_\mu^1W_\nu^1+W_\mu^2W_\nu^2)   \nonumber\\
 &+\Pi_B B_\mu B_\nu + \Pi_1\frac{s_h^2}{4}\left(\frac{g'}{g}B_\mu-W_\mu^3\right)\left(\frac{g'}{g}B_\nu-W_\nu^3\right) \Big], \nonumber \\
\end{align}
where $P_t^{\mu\nu}=g^{\mu\nu}-p^\mu p^\nu/p^2$ and the form factors are
\begin{align}\label{eq:gauge_formfac}
  \Pi_W & =-p^2-\frac{f^2g^2p^2}{m_\rho^2-p^2},\quad \Pi_B=-p^2- \frac{f^2g'^2p^2}{m_\rho^2-p^2},\nonumber\\
  \Pi_1 & =f^2g^2+\frac{2f^2g^2p^2}{m_\rho^2-p^2}-\frac{f^2g^2p^2}{2m_\rho^2-p^2}.
\end{align}
The gauge boson masses can be extracted
\begin{equation}
  m_W=\frac{g f}{2}s_h, \quad  m_Z =\frac{m_W}{\cos\theta_w}.
\end{equation}
The Higgs potential from the effective Lagrangian in Eq.~(\ref{eq:Lg_eff}) after integrating out the gauge bosons in Euclidean momentum space will be
\begin{align}\label{eq:gauge_potential}
  V_g(h)&=\frac{3}{2}\int\frac{d^4p_E}{(2\pi)^4}\left(2\log[\Pi_W+\Pi_1\frac{s_h^2}{4}]\right.\nonumber\\
  &+\left.\log[\Pi_W\Pi_B+\Pi_1\frac{s_h^2}{4}(\Pi_B+\frac{g'^2}{g^2}\Pi_W)]\right).
\end{align}
At order $\mathcal{O}(g^2)$, the gauge contribution to the Higgs potential is
\begin{equation}
  V_g=\gamma_gs_h^2,
\end{equation}
with
\begin{equation}
  \gamma_g=\frac{3}{8(4\pi)^2}\int dp_E^2p_E^2\left(\frac{3\Pi_1}{\Pi_W}+\frac{g'^2\Pi_1}{g^2\Pi_B}\right).
\end{equation}
\section{Fermion Sector}\label{App:top}
In this Appendix we present the fermion sector which  is identical to that in~\cite{Csaki:2018zzf}. The top partner multiplet $\Psi_5$ is in the $\bf 5$ representation of $SO(5)_2$ with mass $M$ introduced at the second site. The left handed top doublet $q_L=(t_L,b_L)$ lives on the first site while the right handed top singlet $t_R$ on the second site, embedded as a singlet of the $SO(5)_2$. The top doublet is embedded in a  $\bf 5$ representation of $SO(5)_1$ with the form
\bea
  \Psi_{q_L}=\frac{1}{\sqrt{2}}\left(
                                 \begin{array}{c}
                                   ib_L \\
                                   b_L \\
                                   it_L \\
                                   -t_L \\
                                   0 \\
                                 \end{array} \right).
\eea
\begin{figure}[t!]
  \centering
  \includegraphics[width=5cm]{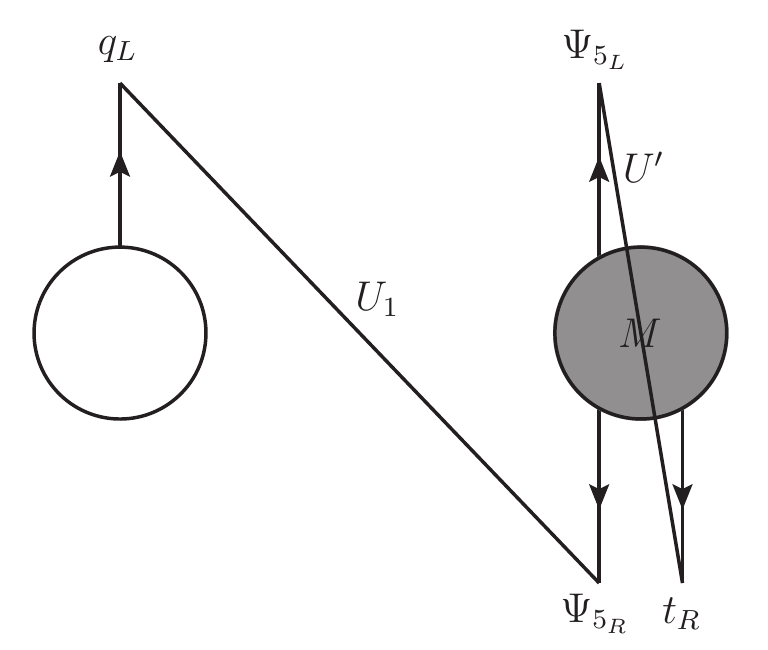}\\
  \caption{Top sector in the two site model with minimal maximal symmetry}\label{fig:2site_top}
\end{figure}

The most general Lagrangian invariant under the gauge symmetries for these fermions is given by
\begin{align} \label{eq:Lag_t}
  \mathcal{L}_t&=\bar{q}_Li\slashed{D}q_L+\bar{t}_Ri\slashed{D}t_R+\bar{\Psi}_5(i\slashed{D}^\prime-M)\Psi_5\nonumber\\
  &-\epsilon_L \bar{\Psi}_{q_L}U_1\Psi_{5R}-\epsilon_R  \bar{\Psi}_{5L} \mathcal{H}' t_R+h.c.\quad.
\end{align}
We find that the top partner sector has an enlarged global $SO(5)_L \times SO(5)_R$ symmetry broken by the mass term to the diagonal subgroup $SO(5)_V$  which is the maximal symmetry. This maximal symmetry will ensure that the effective kinetic terms remain independent of the Higgs and hence result in a finite Higgs potential. After integrating out the Dirac fermion multiplet $\Psi_5$, the effective Lagrangian can be expressed in the global $SO(5)_1$ invariant form in momentum space
\begin{equation}
  \mathcal{L}_{\text{eff}}^t=\Pi_L^0\bar{\Psi}_{q_L}\slashed{p}\Psi_{q_L}+\Pi_R^0\bar{t}_R\slashed{p}t_R+M_t\bar{\Psi}_{q_L}\mathcal{H}t_R+h.c.,
\end{equation}
with the form factors
\bea
  \Pi_L^0  &=&1+\frac{\epsilon_L^2}{M^2-p^2}, \quad  \Pi_R^0= 1+\frac{\epsilon_R^2}{M^2-p^2}, \nonumber \\
  M_t  &=& \frac{M\epsilon_L\epsilon_R}{M^2-p^2}.
\eea
The top mass can be extracted from the above interactions
\bea
  m_t=\frac{\epsilon_L\epsilon_RM}{\sqrt{2}M_TM_{T_1}} s_h,
\eea
where $M_T=\sqrt{\epsilon_L^2+M^2}$ and $M_{T_1}=\sqrt{\epsilon_R^2+M^2}$ are the top partner masses. The Higgs potential from the top sector is of the form
\bea
  V_t(h)=-2N_C\int\frac{d^4p_E}{(2\pi)^4}\log\left(1+\frac{|M_t |^2s_h^2}{2p_E^2\Pi_L^0\Pi_R^0}\right).
\eea
For $s_h \ll 1$, we can expand this potential to $\mathcal{O}(s_h^4)$
\bea
V_t(h) \approx -\gamma_f s_h^2 +\beta_f s_h^4,
\eea
where
\bea\label{eq:gabe_fermion}
\gamma_f &=& \frac{N_C}{(4\pi)^2} \int dp_E^2   \frac{|M_t|^2}{\Pi_L^0 \Pi_R^0},       \nonumber \\
\beta_f &=&\frac{N_C}{(4\pi)^2} \int dp_E^2  p_E^2  \frac{|M_t|^4}{(2p_E^2 \Pi_L^0 \Pi_R^0)^2}.
\eea

We find that $\beta_f$ is quartic in the top Yukawa coupling. According to power counting, $\beta_f$ can be parametrized as
\bea\label{eq:betaf_couting}
\beta_f \approx c_f \frac{N_c y_t^4f^4}{(4\pi)^2 } \ln\frac{M_f^2}{m_t^2},
\eea
where $c_f$ is an ${\cal O}(1)$ numerical constant and $M_f$ is top partner mass scale. $\beta_f$ is  insensitive to the top partner masses. As discussed in~\cite{Csaki:2018zzf}, $\beta_f$ is not big enough to produce the observed physical Higgs mass, hence the  additional adjustable Higgs quartic is necessary and also helpful to reduce the tuning.

\section{the $SO(8)/SO(7)$ twin Higgs model}\label{App:twin_Higgs}
In this final Appendix we summarize the structure of the 2-site $SO(8)/SO(7)$ Twin Higgs model. It is  similar to the 2-site $SO(5)/SO(4)$ model considered above except  one also needs to include the twin sector.
The Lagrangian of the gauge sector has the same form as Eq.(\ref{eq:gauge_Lag}), except that the EW twin gauge fields $\tilde{g}\tilde{A}_\mu^a\tilde{T}_L^a+\tilde{g}'\tilde{B}_\mu\tilde{T}_R^3$ should also be added into  the covariant derivative $D_\mu$. Here $\tilde{T}_L^a$ and $\tilde{T}_R^3$ are the generators of the twin sector's gauge symmetry $SU(2)'_L\times U(1)'_Y$ embedded in the subgroup $SO(4)_2$ which act on the last four indices of $SO(8)_1$. If the gauge couplings are equal, $g=\tilde{g}$ and $g'=\tilde{g}'$, the Lagrangian will have a $Z_2$ exchange symmetry defined as
\begin{equation}
  \mathbb{A}_\mu\leftrightarrow P\tilde{\mathbb{A}}_\mu P,\;U_1\rightarrow PU_1,\;U'\rightarrow U'P_0,
\end{equation}
where $\mathbb{A}_\mu\equiv g A_\mu^aT_L^a+g'B_\mu T_R^3$, $\tilde{\mathbb{A}}_\mu\equiv g\tilde{A}_\mu^a\tilde{T}_L^a+g'\tilde{B}_\mu\tilde{T}_R^3$ and
\begin{equation}
  P=\left(
      \begin{array}{cc}
         & \mathds{1}_4 \\
        \mathds{1}_4 &  \\
      \end{array}
    \right),\;P_0=\left(
                    \begin{array}{cccc}
                       &  & \mathds{1}_3 &  \\
                       & -1 &  &  \\
                      \mathds{1}_3 &  &  &  \\
                       &  &  & 1 \\
                    \end{array}
                  \right).
\end{equation}
It's easy to check that this $Z_2$ is an exchange symmetry between the SM EW sector and its twin sector which include the Higgs TP($A_\mu^a\leftrightarrow \tilde{A}_\mu^a,B_\mu\leftrightarrow \tilde{B}_\mu,s_h\leftrightarrow c_h$). So the Higgs potential must be $Z_2$ invariant.

For the fermion sector, the embedding of the top and twin top into $SO(8)_1$ is
\begin{equation}
 \Psi_{q_L}=\frac{1}{\sqrt{2}}\left(
                                 \begin{array}{c}
                                   ib_L \\
                                   b_L \\
                                   it_L \\
                                   -t_L \\
                                   \mathbf{0}_{4\times 1} \\
                                 \end{array} \right),\;
 \Psi_{\tilde{q}_L}=\frac{1}{\sqrt{2}}\left(
                                 \begin{array}{c}
                                 \mathbf{0}_{4\times 1} \\
                                   i\tilde{b}_L \\
                                   \tilde{b}_L \\
                                   i\tilde{t}_L \\
                                   -\tilde{t}_L \\
                                 \end{array} \right).
\end{equation}
At the second site, an $SU(3)'_c$ triplet $\tilde{\Psi}_8$ in the fundamental representation of $SO(8)_2$ should also be introduced to interact with twin top fields. Similar to Eq.(\ref{eq:Lag_t}),the $Z_2$ invariant Lagrangian is parameterized as
\begin{align}
 \mathcal{L}_f&=\bar{q}_Li\slashed{D}q_L+\bar{t}_Ri\slashed{D}t_R+\bar{\Psi}_8(i\slashed{D}^\prime-M)\Psi_8\nonumber\\
  &-\epsilon_L \bar{\Psi}_{q_L}U_1\Psi_{8R}-\epsilon_R  \bar{\Psi}_{8L} \mathcal{H}' t_R+h.c.\nonumber\\
  &+\big(q_L,t_R,\Psi_8\rightarrow\tilde{q}_L,\tilde{t}_R,\tilde{\Psi}_8\big),
\end{align}
where the $Z_2$ transformation of the fermion multiplets is defined as
\begin{equation}
  \Psi_{q_L}\leftrightarrow P\Psi_{\tilde{q}_L},\;\Psi_8\leftrightarrow\tilde{\Psi}_8,\;t_R\leftrightarrow\tilde{t}_R.
\end{equation}
The Higgs potential from fermion sector will also be $Z_2$ invariant.

\end{document}